  \documentclass[manuscript]{emulateapj}

\usepackage{apjfonts}

\usepackage[usenames]{color}

\slugcomment{to appear in the Astrophysical Journal}

\shorttitle{novae in globular clusters}
\shortauthors{Kato et al.}

\begin{document}


\title{NOVAE IN GLOBULAR CLUSTERS}


\author{Mariko Kato} 
\affil{Department of Astronomy, Keio University, Hiyoshi, Yokohama
  223-8521, Japan;}
\email{mariko@educ.cc.keio.ac.jp}

\author{Izumi Hachisu}
\affil{Department of Earth Science and Astronomy, College of Arts and
Sciences, University of Tokyo, Komaba, Meguro-ku, Tokyo 153-8902, Japan}

\and

\author{Martin Henze}
\affil{European Space Astronomy Centre, P.O. Box 78, 28691 Villanueva de la Ca\~{n}ada, Madrid, Spain}



\begin{abstract}  
We present the first light curve analysis of Population II novae that appeared 
in M31 globular clusters. 
Our light curve models, based on the optically thick wind theory, 
reproduce well both the X-ray turn-on and turnoff times 
with the white dwarf (WD) mass 
of about 1.2 $M_\odot$ for M31N 2007-06b in Bol 111 and   
about 1.37 $M_\odot$ for M31N 2010-10f in Bol 126.   
The transient supersoft X-ray source CXO J004345 in Bol 194 is highly 
likely a nova remnant of 1.2 -- 1.3 $M_\odot$ WD.
These WD masses are quite consistent with the temperatures deduced from X-ray spectra. 
We also present the dependence of nova light curves 
on the metallicity in the range from [Fe/H]=0.4 to $-2.7$.  
Whereas strong optically thick winds are accelerated in Galactic disk novae 
owing to a large Fe opacity peak, only weak winds occur in Population II novae 
with low Fe abundance.  
Thus, nova light curves are systematically slow in low Fe environment. 
For an extremely low Fe abundance normal nova outbursts may not occur unless 
the WD is very massive. 
We encourage $V$ or $y$ filter observation rather than $R$ as well as high cadence 
X-ray monitorings to open quantitative studies of extragalactic novae. 
\end{abstract}

\keywords{ nova, cataclysmic variables   -- stars: individual (M31N 2007-06b, 
M31N 2010-10f, CXO J004345)
-- stars: Population II  -- X-rays: binaries 
}



\section{INTRODUCTION} \label{sec_introduction}

Novae have been observed in the Galaxy, M31, the Large and Small Magellanic Clouds (LMC/SMC), 
and other galaxies 
in various wavelengths from radio to X-ray. 
Many works on extragalactic nova surveys have been reported so far 
\citep[for substantial light curve information, e.g.][and references therein]
{arp56,cao12,cia87,hen10,hen11,hub29,lee12,nei04,ori06,ori10,pie05,ros64,ros73,sha01,sha11a,sha11b,sha12,sha13a}. 
Different trends have been suggested in the distribution of nova speed classes 
and light curve types among the Galaxy, M31 and LMC, as well as for the Galactic disk and bulge
\citep{pay57,due90}.
Recent surveys of M31 found a spatial difference between fast and slow novae;  
slower novae, or less massive white dwarfs (WDs), tend to concentrate towards the inner part 
of M31 while fast novae (or massive WDs) are distributed in a wider region \citep{hen11,sha11b}. 
LMC novae tend to have faster declines \citep[e.g.][]{due90,sha13a}. 
Novae that have appeared in globular clusters were all fast (see Section \ref{sec_GCnova} for
summary).

These differences are sometimes discussed in terms of 'different stellar populations.' 
\citet{sha13a} interpreted the difference in the speed class and spectral type 
between galaxies as the difference in the dominant stellar populations as follows. 
A significant fraction of LMC novae belong to a younger population that has   
a shorter elapsed time from its birth. Younger stars on average produce more massive WDs  
than older stars do. More massive WDs lead to faster nova outbursts. 
Therefore, LMC novae are on average faster than M31 novae.

We note, however, that besides the classification of stars as 
belonging to young or old stellar populations
there is also the distinction between Population I/II stars.
These two are not always the same. Populations I and II indicate 
a difference in metallicity (in particular [Fe/H]) whereas young/old designate a 
distinction in time since the stellar birth. 
Differences in the nova speed class may reflect not only different WD masses but also 
different metallicities.

Theoretically, the nova evolution can be explained by the optically thick wind theory.
The mass ejection and nova timescales are governed by continuum-driven winds which
are accelerated owing to an iron opacity peak deep inside the photosphere
\citep[e.g.][]{kat94h}.  
In general, massive WDs show faster declines because of strong accelerations of winds
\citep{hac06}. 
\citet{kat97} calculated theoretical light curve models of novae with different
metallicities and showed that nova light curves depend strongly on the iron content.
For the same WD mass, the nova evolution is affected by metallicity, especially by iron content,
and low metallicity novae evolve much slower.
This may explain the different trends between disk novae and bulge novae.
As bulge stars on average have a lower metallicity than disk stars, we expect that
nova outbursts are slower in the bulge even if the WD mass distributions are the same.
Bulge stars show, however, non-uniform metallicities, both in the Galaxy and in M31,
from $Z/Z_\odot$ =0.2 to 2 \citep{sar05}.
Therefore, we need to know both the metallicity and WD mass of individual novae
in order to clarify the physical reason for the different observed properties.
This is, however, difficult for normal field stars.

The present work concentrates on novae that appeared in M31 globular clusters
to determine the WD mass from light curve analysis.
Compared with studies of Galactic novae, M31 globular cluster novae have merits
such as that (1) the distance is known,
(2) [Fe/H] is already estimated for each cluster, and
(3) internal extinction within a cluster can be neglected.
Therefore, we can solve the degeneracy of two effects, the WD mass and metallicity.
This is the first step to our quantitative study of extragalactic novae
with different metallicity environments.

The next section (Section \ref{sec_galacticnova}) 
gives a brief summary of nova speed classes and light curves of typical Galactic novae. 
Section \ref{sec_theory} shows theoretical calculations of nova envelopes  
with various [Fe/H].  Section \ref{sec_GCnova} describes the light curve analysis for  
two novae and one nova candidate in M31 globular clusters. 
Section \ref{sec_discussion} raises some issues on optical 
and X-ray surveys. Conclusions follow in Section \ref{sec_conclusions}.


\begin{figure*}
\epsscale{0.8}
\plotone{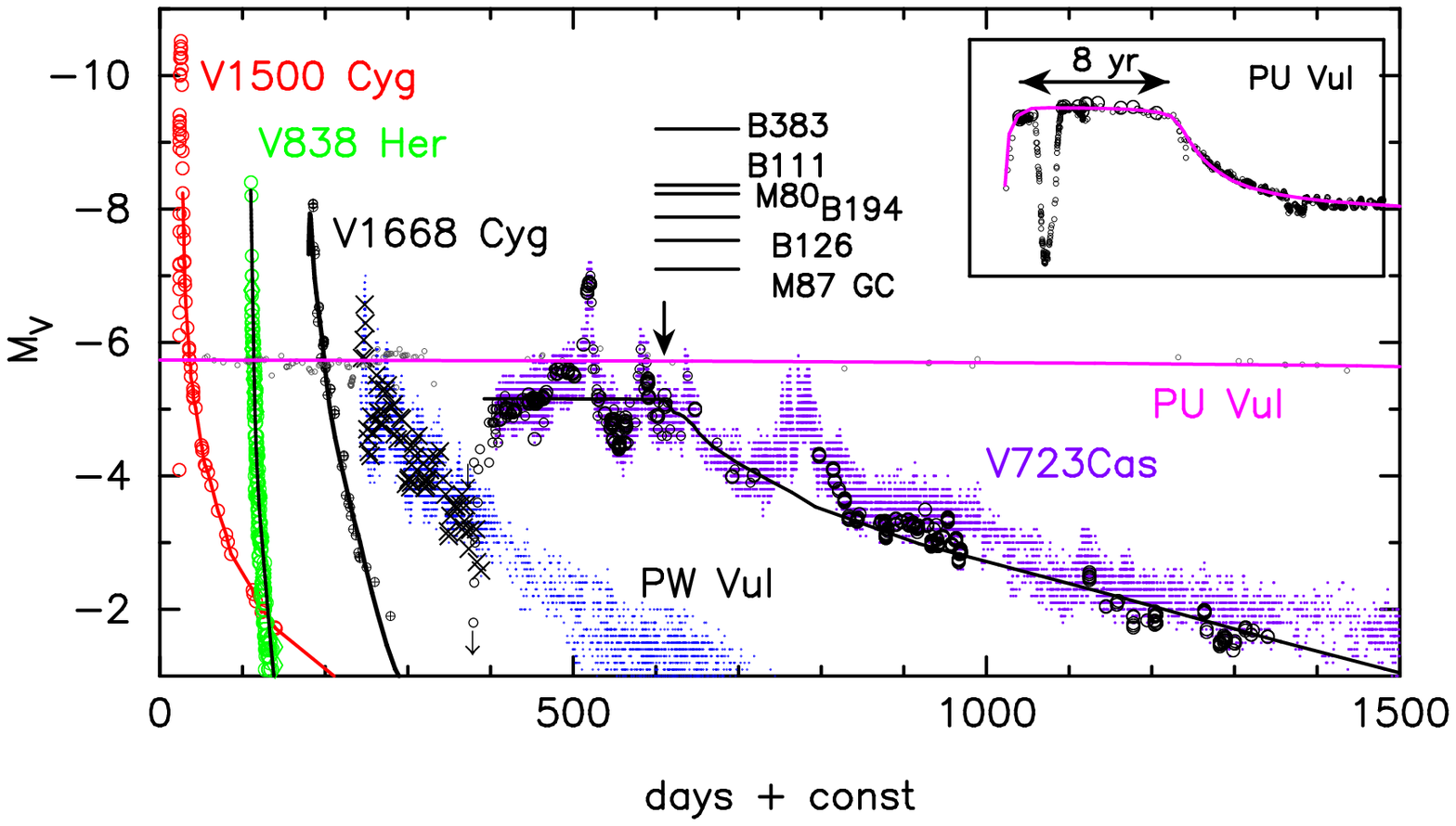}
\caption{Light curves of various speed classes of Galactic classical novae.
The very fast nova V1500 Cyg is a superbright nova (red) \citep[data taken from][]{tem79}.
V838 Her (green) is a very fast nova \citep[data taken from][]{kat09}.
V1668 Cyg (black) is a typical fast nova \citep[data taken from][and AAVSO]{gal80}.
PW Vul (blue) is a slow nova \citep[data taken from][:X-marks and AAVSO: dots]{rob95}.
V723 Cas (purple and black) is a very slow nova
\citep[data taken from][and AAVSO and VSNET]{cho97,cho98}.
PU Vul (gray horizontal line) is a symbiotic nova and only the flat optical peak
is plotted. The inset shows the first 20 yr of its light curve \citep{kat12hm}.
The solid lines running through each nova indicate theoretical light curve models.
See Section \ref{sec_nova_summary} for the WD mass and the chemical 
composition of these theoretical models.
The downward arrow indicates the assumed transition point of V723 Cas
from static to wind envelope structure \citep{kat11h}. 
Brightness of six globular clusters hosting novae are indicated by black solid 
short horizontal lines. 
\label{lightcurve}
}
\end{figure*}

\section{ABSOLUTE BRIGHTNESS OF GALACTIC NOVAE} \label{sec_galacticnova}

Figure \ref{lightcurve} shows $V$ light curves of six well-observed
Galactic classical novae with different speed classes.
V1500 Cyg is an exceptionally bright, very fast nova ($t_2=2.9$ and $t_3=3.6$ days);
a so-called superbright nova \citep{del91}.
Here, $t_2$ ($t_3$) is the time in days in which a nova decays by two (three) 
magnitudes from maximum.
V838 Her is a very fast nova, one of the fastest novae except the superbright novae.
It shows a normal super Eddington luminosity in the early phase.
V1668 Cyg is a fast nova of $t_2 =12$ and $t_3=25$ days.
PW Vul is a slow nova that shows oscillatory behavior in the early phase.
V723 Cas is a very slow nova that features multiple peaks in the early stage.
PU Vul is an extremely slow symbiotic nova with a flat optical
peak lasting eight years. The first 20 years of its light curve appear in the inset of Figure \ref{lightcurve}.

In this work, we adopt the classification scheme of nova speed class after \citet{pay57}. 
(1) very fast: $t_2 <$ 10 days;  
(2) fast: $t_2 =$ 11 -- 25 days;
(3) moderately fast: $t_2 =$ 26 -- 80 days;
(4) slow:  $t_2 =$ 81 -- 150 days;
(5) very slow: $t_2 =$ 151 -- 250 days.

The absolute magnitude is obtained from the distance modulus, 
$(m-M)_{\rm V} = 5 \log(d/10) + A_{\rm V}$, 
as follows: 
$(m-M)_V =12.3$ with $E(B-V)=0.45$ for V1500 Cyg, 
$(m-M)_V =12.2$ with $E(B-V)=0.30$ for V1974 Cyg, 
$(m-M)_V =14.25$ with $E(B-V)=0.35$ for V1668 Cyg, 
$(m-M)_V =13.0$ with $E(B-V)=0.55$ for PW Vul,
$(m-M)_V =14.0$ with $E(B-V)=0.35$ for V723 Cas, 
$(m-M)_V =14.3$ with $E(B-V)=0.30$ for PU Vul, 
$(m-M)_V =13.9$ with $E(B-V)=0.70$ for V5558 Sgr  
\citep{hac13} 
and $(m-M)_V =13.8$ with $E(B-V)=0.53$ for V838 Her \citep{kat09}. 
We adopt $A_V =3.1 E(B-V)$. 

The solid lines superposed on the data represent theoretical light curve models which are
fitted with multiwavelength observational data including
IR, optical, UV, and supersoft X-ray.
In V1500 Cyg, the WD mass is estimated to be $M_{\rm WD}=1.20~M_\odot$
for $X=0.55,~Y=0.3,~X_{\rm CNO}=0.1$, $X_{\rm Ne}=0.03$, and $Z$=0.02 \citep{hac13}.
In V838 Her, $M_{\rm WD}=1.35~M_\odot$ for $X=0.55,~Y=0.33,~X_{\rm O}=0.03$,
$X_{\rm Ne}=0.07$, and $Z$=0.02 \citep{kat09}.
For V1668 Cyg, $M_{\rm WD}=0.95~M_\odot$ with
$X=0.45,~Y=0.18,~X_{\rm CNO}=0.35$ and $Z$=0.02 \citep{hac06}.
In V723 Cas, 0.6 $M_\odot$ for solar composition \citep{kat11h}.
In PU Vul, 0.6 $M_\odot$ for $X=0.55,~Y=0.33$ and $Z=0.02$ \citep{kat12hm}.
The precision of these WD mass estimates is about $\pm 0.05 M_\odot$.
As shown in Figure \ref{lightcurve}, the speed class and peak brightness roughly
reflect the WD mass except for the superbright nova V1500 Cyg.
The faster and brighter the nova, the larger the WD mass.

Figure \ref{lightcurve} also shows the brightnesses of six
globular clusters hosting novae as shown later in Section \ref{sec_GCnova}.
The absolute $V$ magnitudes of M31 globular clusters range widely from $-6$ mag to
$-11$ mag \citep{van10}.
In extragalactic globular clusters we cannot resolve a nova from
stars in the hosting cluster with small ground based telescopes usually used
in nova surveys. Therefore,
if these Galactic novae would appear in a bright extragalactic globular cluster,
we are likely to detect only fast novae like V838 Her and V1668 Cyg
but could miss slower novae like PU Vul and V723 Cas.
In a fainter cluster, with $M_V \sim -6$, we may
detect most of such novae. However, a fainter cluster contains a small population of
stars and thus small probability of nova outbursts during a nova survey.
We think that Bol 126 (Section \ref{sec_M31N2010-10f}) is a reasonable size of cluster,
bright enough to host a sufficiently large population of stars
but faint enough for a fast nova emerging from the host cluster to be detected.


\begin{figure*}
\epsscale{0.8}
\plotone{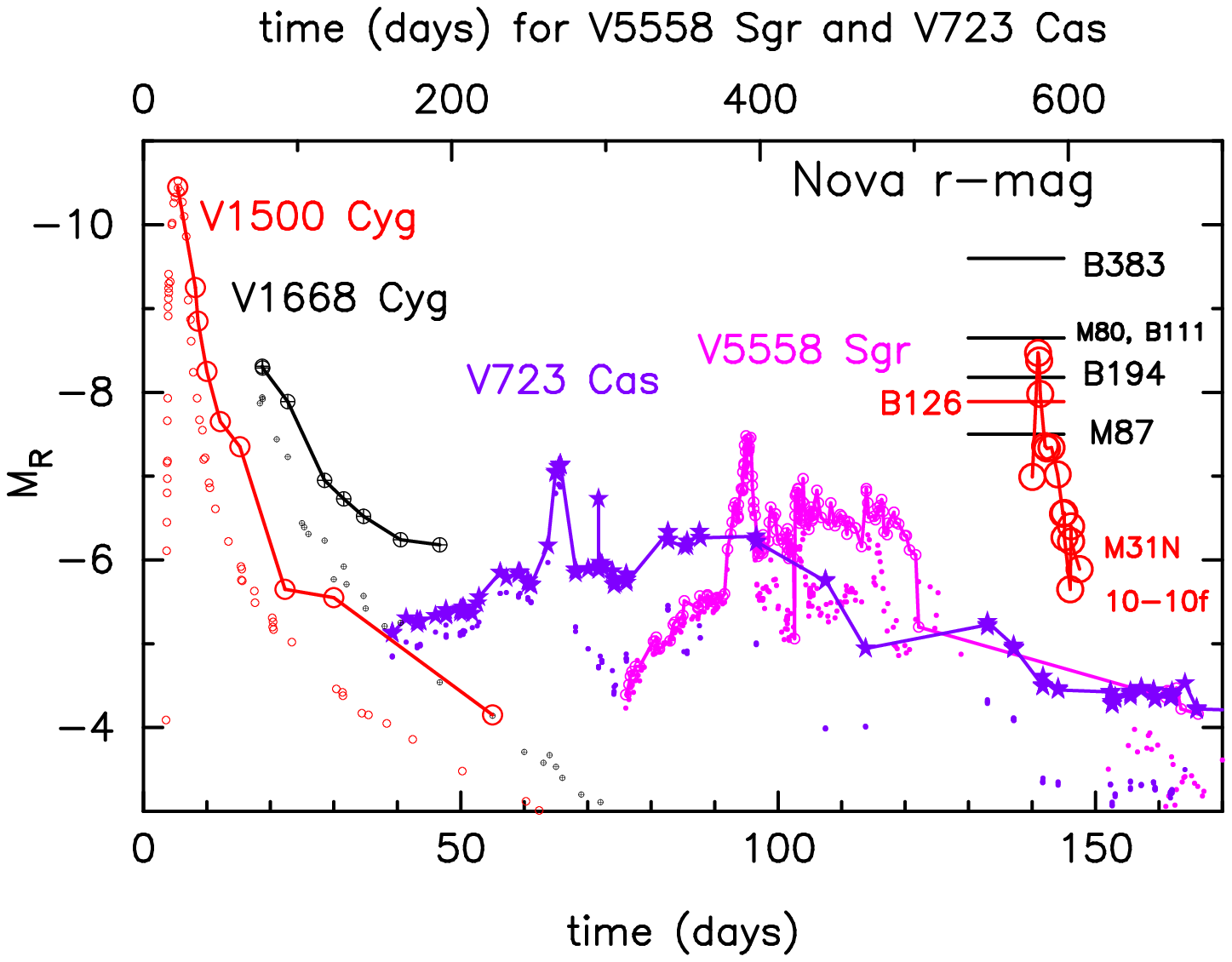}
\caption{Nova light curves in R-magnitude.
The light curve of M31N 2010-10f
\citep{hen13} is plotted by red large open circles as well as the brightness of its
host globular cluster Bol 126.
The $R$-magnitude light curves of Galactic novae are plotted by lines with various symbols:
the very fast nova V1500 Cyg \citep{gal76},
the fast nova V1668 Cyg \citep{der78}
and two slow novae V723 Cas \citep{cho97,cho98} and V5558 Sgr (AAVSO).
The smaller symbols of the same color below each $R$-magnitude light curve 
indicate the $V$ magnitude data appeared in Fig. \ref{lightcurve}.
The timescale for two slow novae is squeezed by a factor of $\sim 4.1$ in the upper abscissa.
\label{nova.rmag}
}
\end{figure*}

Figure \ref{nova.rmag} shows $R$ band light curves of four Galactic novae,
V1500 Cyg, V1668 Cyg, V723 Cas, and V5558 Sgr.
Each light curve consists of $R$ band data
(the upper larger symbols connected by a solid line)
and optical data ($V$/visual/$y$: the lower smaller symbols of the same color).
Here, we adopt $A_R = 2.33 E(B-V)$ \citep{car89}.
PU Vul is not shown in this figure in order to avoid complexity of lines,
but it is slightly brighter in the $R$ band (from $-6.1$ to $-6.3$ mag) than in the $V$ band
in the later part of the flat maximum \citep{shu12}.

A nova outburst is brighter in the $R$ band than in the $V$ band
because of the contribution of emission lines such as H$_\alpha$ to the broad band brightness.
As shown in Figure \ref{nova.rmag}, the $R$ magnitudes of V1500 Cyg and V1668 Cyg
are always brighter than the respective $V$ magnitudes, but the
difference between them is small around the peak and increases as they decay.
This tendency is also seen in the multi-peak novae V5558 Sgr (Figure \ref{nova.rmag}),  
V458 Vul, and V378 Ser \citep{tan11b} 
and the single re-brightening nova V2362 Cyg \citep{mun08}
for which the difference is smaller around the peaks than
during the inter-peak phase. 
The multi-peak slow novae V2540 Oph, V4745 Sgr, and V458 Vul showed in their spectra that   
H$_\alpha$ emission line was narrower (with small equivalent width) 
and weaker (relative to the continuum) near the maxima but wider and stronger 
in the inter-peak phase (private communication with D. Nogami, 2013).  

We also added the $R$ light curve of nova M31N 2010-10f which appeared in Bol 126
(see Section \ref{sec_M31N2010-10f} for details).
Its peak luminosity is comparable to that of V1668 Cyg, a typical Galactic
fast nova, but it shows a much faster decay.
\citet{hen13} estimated this $R$ light curve until when the nova had decayed more than 2 magnitudes
from the peak by subtracting the original globular cluster brightness from the observed flux.
In this way, we can possibly detect novae of 1--2 magnitude fainter than
the host cluster. We will discuss the detectability in more detail
in Section \ref{sec_discussion}.

\section{POPULATION II NOVAE -- THEORY} \label{sec_theory}

\subsection{Numerical Method} \label{sec_model}

A nova is a thermonuclear runaway event triggered by unstable hydrogen nuclear burning
in the hydrogen-rich envelope on a WD. After the onset of a nova outburst,
its envelope expands to a giant size and the optical magnitude
reaches the maximum. Strong winds are continuously accelerated,
and carry most of the envelope matter away from the binary.

The decay phases of novae have been extensively calculated
by Kato and Hachisu and their collaborators for Galactic novae.
The optical and infrared light curves
are well described by the free-free emission model \citep[e.g.][]{hac06} based on the optically
thick wind theory \citep{kat94h}.
Nova light curves depend strongly on the WD mass and weakly
on the chemical composition if we fix the iron content.
Novae on more massive WDs evolve faster than on less massive WDs.
\citet{hac06} found a homologous shape of nova light curves
independent of the WD mass and chemical composition, i.e.,
if we normalize each nova time-scale,
all the optical/IR light curves in the decay phase become very similar.
This property is theoretically explained in terms of free-free emission based on
the optically thick wind theory by
\citet{hac06}, who dubbed this property the universal decline law of classical novae.

The dependence of nova light curves on the stellar
population (i.e., iron abundance) was studied
by \citet{kat97} from $Z=0.5$ to $Z=0.001$. Here, $Z$ means the so-called
heavy element content by weight of accreted matter, not of nova ejecta.
Population II novae evolve more slowly than Population I novae, because
the acceleration of optically thick winds is continuum radiation-driven
in the Fe ionization region where the opacity has its largest peak \citep{kat97}.
We extend the models of \citet{kat97} down to
$Z=0.0001$, i.e., extreme Population II in order to obtain a more complete picture of
nova behavior including very low Fe abundance.


\begin{deluxetable*}{llllllccc}
\tabletypesize{\scriptsize}
\tablecaption{Novae and A Nova Candidate in Globular Clusters
\label{table_nova}}
\tablewidth{0pt}
\tablehead{
\colhead{Object} &
\colhead{  } &
\colhead{Nova/SSS} &
\colhead{Year} &
\colhead{Galaxy}&
\colhead{Globular}&
\colhead{[Fe/H]\tablenotemark{a}} &
\colhead{$M_{\rm V}$(GC)\tablenotemark{b}}&
\colhead{$M_{\rm R}$(GC)\tablenotemark{c}}\\
\colhead{  } &
\colhead{  } &
\colhead{  } &
\colhead{  } &
\colhead{  } &
\colhead{cluster } &
\colhead{  } &
\colhead{  }
}
\startdata
M31N 2007-06b & ... & fast nova/SSS&2007  & M31& Bol 111  & --1.3&--8.36& --8.65 \\
M31N 2010-10f & ... & very fast nova/SSS&2010  &M31& Bol 126 &--1.5&--7.53 &--7.89 \\
CXO J004345\tablenotemark{d}& ...& ---/SSS & 2007 &M31& Bol 194& --1.40&--7.88&--8.18\\
Anonymous & ... & possible nova/---&1991   &M31& Bol 383 & --0.47,--0.6 &--9.2 &--9.6 \\
T Sco    &      ... & fast nova/---&1860 &Galaxy&M80 (NGC6093) &--1.75&--8.23& --8.65\\
N2001 in M87  & ... & fast nova/---&2001    & M87 &anonymous& \nodata& --7.1& --7.5
\enddata
\tablenotetext{a}{Bol 111, Bol 126 and Bol 194: \citet{cal11}, 
Bol 383: \citet{gal09,cal11},
M80: \citet{van10}.}
\tablenotetext{b}{B111, B126, B194: see the text in Section \ref{sec_GCnova}. 
B383: \citet{gal04}. M80:\citet{har10}. M87 GC: \citet{sha04}.}
\tablenotetext{c}{B111, B126, B194: Section \ref{sec_GCnova}.  B383: \citet{gal04}. 
     M80: \citet{har10}. M87 GC: from $I=-8.06$ \citep{sha04} with assumed $I-R=-0.53$ of M31 GCs' mean.}
\tablenotetext{d}{We regard SS2 in \citet{hen09} as CXO J004345.4+410611.}
\end{deluxetable*}

We assume that the chemical composition of the envelope is uniform. We adopt
$X=0.55$, $X_{\rm C}=0.1$, $X_{\rm O}=0.1$,  and $Z=0.05,~0.02,~0.004,~0.001,~0.0004$, and
0.0001 for CO nova ($0.5 M_\odot \leq M_{\rm WD} < 1.1 M_\odot$).
Here, $X_{\rm C}$ and $X_{\rm O}$ denote the mass fraction of carbon and oxygen.
Note that $Z$ includes carbon and oxygen
as well as other heavy elements.

For less massive WDs ($M_{\rm WD} < 0.5 M_\odot$)
we suppose that the underlying WD is a helium WD and adopt a
helium-rich composition as $X$=0.5 and $Y=1-X-Z$.
For massive WDs ($ M_{\rm WD} \geq 1.1 M_\odot$) we adopt a
composition for neon novae as $X=0.5$, $X_{\rm O}=0.1$, $X_{\rm Ne}=0.1$.
Here, $X_{\rm Ne}$ is the mass fraction of neon.
For comparison, we also did calculations for other sets of compositions,
which will be described separately.
We use the OPAL opacity tables \citep{igl96} with a mixture of $Z$ ingredients
of solar composition (Grevesse and Noels (1993) mixture)
for Population I stars ($Z \geq 0.001$) and alpha-element enhancement
composition (Weiss (1995) mixture) for Population II stars ($Z = 0.0001$ and 0.0004).
We adopted $X=0.55$ instead of $X=0.35$ \citep{kat97}, based on the experience of many
light curve fittings of Galactic novae that support a composition
of $X=0.55$ rather than of $X=0.35$ in average.
Our numerical method, basic equations, and boundary conditions are all the same as those
in the previous paper \citep{kat94h}. Optical light curves are calculated based on free-free
emission, when the optically thick winds occur, and blackbody photospheric emission, when
the wind is not accelerated (in low mass WDs). X-ray light curves are calculated
for the energy range between 0.2 and 1.0 keV,
assuming a blackbody with the photospheric temperature of the envelope solution.


\begin{figure}
\epsscale{1.1}
\plotone{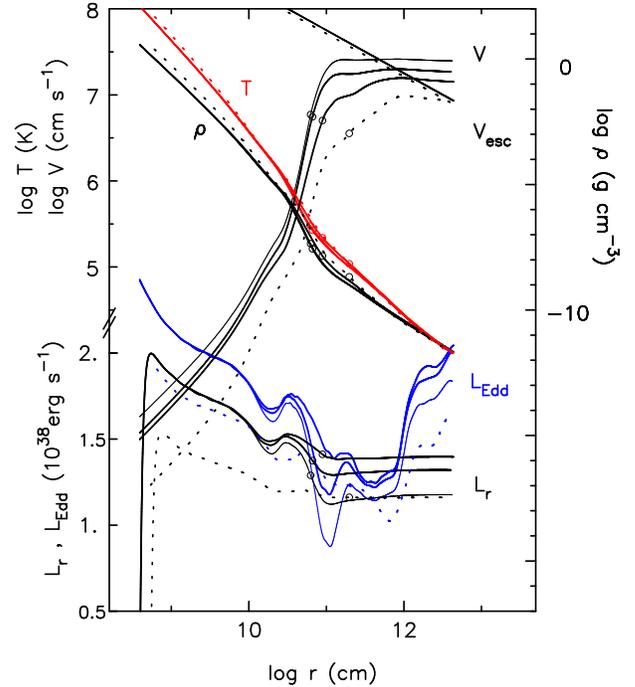}
\caption{Structures of the envelope on a $1.2 M_\odot$ WD
($X=0.55, ~X_{\rm O}=0.1,~X_{\rm Ne}=0.1$) with different $Z$ but
the same photospheric temperature of $\log T_{\rm ph}$(K)=4.0.
From upper to lower in the velocity $v$ (lower to upper for the diffusive
luminosity $L_r$ and the local Eddington luminosity
$L_{\rm Edd}=4 \pi c G M_{\rm WD}/ \kappa$),
$Z=0.004$ ($R_{\rm cr}=0.9~R_\odot$, thin solid line),
$Z=0.001$ ($R_{\rm cr}=0.95~R_\odot$ thick solid line in the middle),
$Z=0.0004$ ($R_{\rm cr}=1.28~R_\odot$ thick solid line).
The position of the critical point $R_{\rm cr}$ is indicated by a small open circle.
Temperature $T$ and density $\rho$ profiles almost overlap.
The wind velocity finally exceeds the escape velocity $V_{\rm esc}$.
The dotted lines show another example of weak winds: a 1.0 $M_\odot$ WD model
with $\log T_{\rm ph}$(K)=4.0 ($R_{\rm cr}=2.83~R_\odot$) (see Section \ref{sec_speedclass}).
\label{struc}
}
\end{figure}

\subsection{Envelope Solutions: Weak Acceleration of Winds}\label{sec_speedclass}

Figure \ref{struc} shows envelope structures on a $1.2~M_\odot$ WD
with $X=0.55$, $X_{\rm O}=0.1$, $X_{\rm Ne}=0.1$, and $Z=0.0004,~0.001$ and 0.004 with the
same photospheric temperature of $\log T_{\rm ph}$(K)$=4.0$. The internal structures
in the temperature $T$ and density $\rho$ are almost the same, but
those in the velocity $V$ and luminosity $L_r$ profiles are clearly different.
The diffusive luminosity decreases outward on the
outside of the nuclear burning region ($\log r$(cm)$ \sim 8.59-8.73$)
because a part of the radiation energy is consumed to accelerate the winds.
The wind is strongly accelerated deep inside the envelope
and the velocity reaches the terminal
velocity below the photosphere. The critical point of solar-wind type
solution \citep[for the exact definition, see][]{kat94h} appears
in the acceleration region, which is depicted by small
open circles. The local Eddington luminosity
 $L_{\rm Edd}=4 \pi c G M_{\rm WD}/ \kappa$ has a minimum corresponding
to each opacity peak.


\begin{figure}
\epsscale{1.1}
\plotone{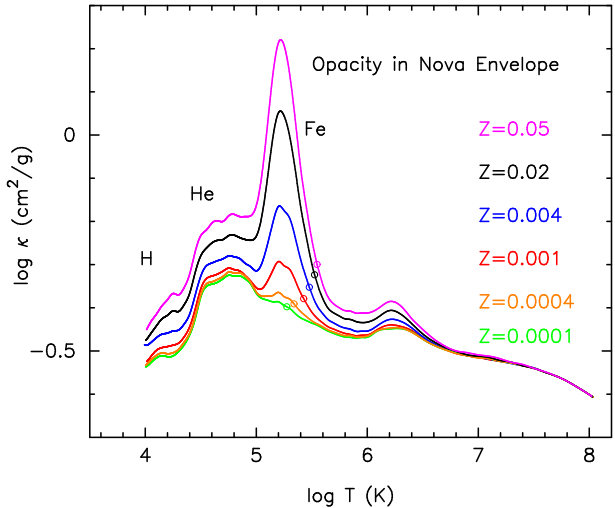}
\caption{Run of the OPAL opacity \citep{igl96} for the wind solutions of
$1.2 M_\odot$ WD envelope with $\log T_{\rm ph}=4.0$.
The chemical composition is assumed to be $X=0.55$, $X_{\rm O}=0.1,~X_{\rm Ne}=0.1$, and
$Z=$0.05, 0.02, 0.004, 0.001, 0.0004, 0.00001 from upper to lower.
Small open circles denote the critical points.
The main element responsible for each opacity peak is indicated by its atomic symbol
(see the text for more details).
\label{opac}
}
\end{figure}

Figure \ref{opac} shows the run of the OPAL opacities of $1.2~M_\odot$ WD
envelopes with $\log T_{\rm ph}$ (K) =4.0 including the same three models as
in Figure \ref{struc}.
The opacity has several peaks above the constant value of
the electron scattering opacity $\log \kappa_{\rm el} =\log [0.2(1+X)]=
-0.51$ with $X=0.55$.
The small peak at
$\log T$(K) $\sim 4.1$ is due to hydrogen ionization.
The second peak from the left at $\log T$ (K) =4.5 -- 4.8 corresponds
to helium ionization.
The prominent peak at $\log T$ (K) $\sim 5.2$ is owing mainly to
low/mid-degree ionized iron.
This iron contribution was found in opacity projects
\citep{igl87,sea94}.
The peak at $\log T$ (K)= 6.2 -- 6.3 relates to highly ionized iron, O, and Ne.
There is also a tiny peak at/around $\log T$ (K)= 7.1 from the highly ionized
heavy elements Ar -- Fe. The opacity is smaller than the value of electron
scattering at the highest temperature region because of the Compton effect.
In low-$Z$ models the Fe peak at $\log T$ (K) $\sim 5.2$ is as low as
the He peak and almost disappears in $Z < 0.001$ models.

Figures \ref{struc} and \ref{opac} show
the critical point of each model (small open circles) that shifts outward
as $Z$ decreases, but still stays inside the photosphere.
In high $Z$ models, the strongest Fe peak mainly drives the
winds and the critical point appears just inside the Fe peak.
In contrast, for very low $Z$ models ($Z=0.0001$ and 0.0004)
the Fe peak is small and the He peak contributes to the acceleration of the winds.
However, in this case, the acceleration is very weak and
the velocity either barely reaches the escape velocity near the photosphere or
does not reach it at all. This tendency is already reported by \citet{kat97}, but
our models show this trend more clearly because of weaker acceleration in
lower $Z$ ($<~0.001$) and higher $X$ (=0.55).
In the intermediate $Z$, the two peaks of Fe and He are comparable, which causes
deceleration of winds between the two peaks and in some cases possibly results in
non-smooth light curves with unsteady mass-ejection.
Figures \ref{struc} also shows a solution for a 1.0$M_\odot$ WD envelope,
which we discuss in Section \ref{sec_hewind}.

\subsection{Dependence of Nova Speed Class on Metallicity}\label{theory_Zdepencence}

After the optical maximum, strong winds carry a large part of the envelope mass away and the
photospheric radius decreases with time. Accordingly,
the photospheric temperature rises with time, and the
main emitting wavelength region shifts from optical to UV and finally
to supersoft X-ray.


\begin{figure}
\epsscale{1.15}
\plotone{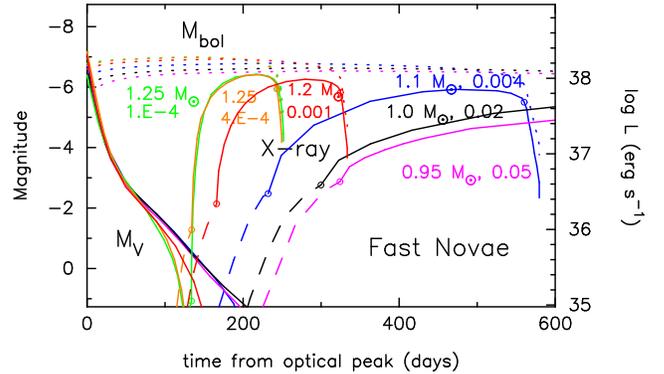}
\caption{Theoretical light curves of the optical free-free ($M_V$: solid line),
bolometric magnitude ($M_{\rm bol}$: dotted), and
supersoft X-ray flux ($0.2-1.0$ keV: 12.4 \AA~-- 62 \AA) for six models.
Their WD masses and chemical compositions of the envelopes are different to each other,
but they have very similar early optical declines.
Their ($M_{\rm WD},~Z$) are as follows:
Green line: ($1.25 M_\odot$, 0.0001).
Orange line: ($1.25 M_\odot$, 0.0004).
Red line: ($1.2 M_\odot$, 0.001).
Blue line: ($1.1 M_\odot$, 0.004).
Black line: ($1.0 M_\odot$, 0.02).
Purple line: ($0.95 M_\odot$, 0.05).
While the optical light curves are very similar in the first 50 days
the X-ray turn-on (small open circle at the boundary of dashed and
solid part) and turnoff times (small open circle at the decreasing
phase of supersoft X-ray flux) are very different.
The dashed part in the X-ray light curve represents the wind phase, in which the
supersoft X-rays may be partly obscured owing to self absorption of the winds.
\label{fast}
}
\end{figure}

Figure \ref{fast} depicts six theoretical models of fast novae that show similar
free-free optical light curves.
These models are ($M_{\rm WD}$, $Z$)=($0.95 M_\odot$, 0.05), ($1.0 M_\odot$, 0.02), 
($1.1 M_\odot$, 0.004), ($1.2 M_\odot$, 0.001), ($1.25 M_\odot$, 0.0004), and
($1.25 M_\odot$, 0.0001).
Here, we assume the standard set of chemical composition as described in Section \ref{sec_model}.
While these models show a similar optical decline
until $M_V = -2$, their X-ray turn-on and turnoff times are very different. 
In general, smaller $Z$ causes weaker wind-acceleration and results in
a slower nova evolution in the early phase of the outburst.
Therefore, a similar optical decline rate is obtained in a more massive WD
with a smaller $Z$. However, a massive WD has a smaller envelope mass
and the winds blow it off in a shorter time.
Thus, we get smaller X-ray turn-on/turnoff times. 


\begin{figure}
\epsscale{1.15}
\plotone{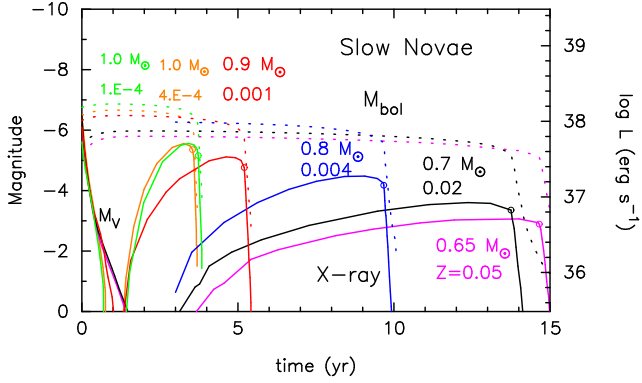}
\caption{Same as Figure \ref{fast}, but for slow novae.
The model parameters of ($M_{\rm WD},~Z$) are as follows:
Green line: ($1.0 M_\odot$, 0.0001). 
Orange line: ($1.0 M_\odot$, 0.0004). 
Red line: ($0.9 M_\odot$, 0.001). 
Blue line: ($0.8 M_\odot$, 0.004). 
Black line: ($0.7 M_\odot$, 0.02). 
Purple line: ($0.65 M_\odot$, 0.05). Since we could not calculate 
the wind phase of the $0.8 M_\odot$ model, we do not plot the optical light curve
and assume the beginning epoch of the X-ray phase arbitrarily.
\label{slow}
}
\end{figure}

Figure \ref{slow} shows light curves of slow novae.
These models are ($M_{\rm WD}$, $Z$)=($0.65 M_\odot$, 0.05), ($0.7 M_\odot$, 0.02),
($0.8 M_\odot$, 0.004), ($0.9 M_\odot$, 0.001), ($1.0 M_\odot$, 0.0004), and
($1.0 M_\odot$, 0.0001).
In all the cases, the combination of a more massive WD with a smaller $Z$ yields a
similar optical decline. Novae enter the supersoft X-ray phase much earlier
in smaller $Z$ models.
In Figures \ref{fast} and \ref{slow}, the models of $Z=0.0004$ and $0.0001$
show very similar results. The difference in $Z$ has nothing to do with the wind acceleration.
This is because the iron opacity peak of these two models
is too small to accelerate winds compared with the helium peak
as in Figure \ref{opac}.


\begin{figure}
\epsscale{1.15}
\plotone{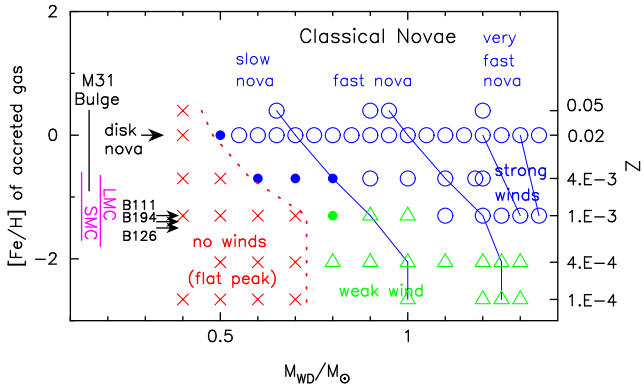}
\caption{Three different regions of wind acceleration in the WD mass vs. metallicity diagram.
[Fe/H] means $\log [(X_{\rm Fe}/X)/(X_{{\rm Fe}\odot}/X_\odot)]$, where we assume 
$X_{{\rm Fe}\odot}=1.436 \times 10^{-3}$ and $X_\odot=0.7$. 
The right ordinate shows $Z$ corresponding to [Fe/H].
Blue open circles: strong winds driven by the Fe opacity peak at $\log T$(K)=5.2.
Green open triangles: weak winds driven by the He opacity peak.
Red X-marks: no winds.
Blue filled circles: wind solutions are not obtained due to numerical difficulty.
The ranges of metallicity in the M31 bulge \citep{sar05} and LMC and SMC \citep{chi09} are
denoted by vertical lines.
Three globular clusters hosting novae  are also shown.
\label{MwdZ}
}
\end{figure}

Figure \ref{MwdZ} summarizes our model calculations in the $M_{\rm WD}$ -- $Z$
diagram (WD mass versus metallicity).
The area denoted by blue open circles corresponds to the region of
strong optically thick winds driven by the Fe opacity peak.
This area corresponds to normal nova outbursts.
In the area denoted by red crosses, no optically thick winds
are accelerated, so novae evolve very slowly. If the extended
stage lasts much longer than 8 years of PU Vul,
we may not recognize the outburst as a nova, but may mis-categorize it
as an F-supergiant.
The filled circles near the boundary of optically thick winds and no-winds
denote the models in which our numerical calculation does not converge
because the envelope model is unstable between the two opacity peaks.
In this area acceleration is too weak and we may not expect a strong
nova outburst with smooth decline in its light curve.
In the area of open triangles, the small opacity peak of He ionization accelerates
weak winds. No novae have been found so far corresponding to this area.
We will discuss on this area in Section \ref{sec_discussion}.

We connect models that have a similar optical decline in Figure \ref{MwdZ}.
In addition to the two sets of light curve models in Figures \ref{fast} and \ref{slow},
we added two pairs of ($M_{\rm WD}$, $Z$)=($1.2 M_\odot$, 0.02) -- ($1.3 M_\odot$, 0.001),
and ($1.3 M_\odot$, 0.02) -- ($1.35 M_\odot$, 0.001).
These four lines demonstrate that low-metallicity, fast novae appear only in a limited range
of very large WD masses.
In low $Z$ ($Z \lesssim 0.0004$) the results are independent of metallicity
because the Fe peak is too small to affect the nova evolutions.

This figure also shows typical values of [Fe/H] in the M31 bulge, LMC, and SMC as well as
three globular clusters which also appear in Table \ref{table_nova}.
A majority of Galactic novae belong to the disk population that is represented by the arrow at $Z=0.02$.
The M31 bulge stars show various metal contents: 60\% of stars show subsolar abundance
and a quarter of stars shows $Z \lesssim 0.008$ \citep{sar05}.
Thus, we guess that a large part of the M31 bulge novae correspond to a region slightly
lower than $Z=0.02$. The Galactic bulge stars show
similar metal distribution to that of M31 bulge stars,
with slight shift to lower metal contents \citep{sar05}.
Thus, we expect a systematic difference in the speed class of novae
between the Galactic disk and Galactic/M31 bulge.
If the mass distributions of WDs are the same,
we expect that bulge novae tend to be slower compared with Galactic disk novae
because of smaller [Fe/H] in average.
In environments with much lower metal content ($Z \sim 0.001$), such as globular clusters, the LMC or SMC,
we can expect fast novae on massive WDs while slow novae are uncertain to be realized
(see Section \ref{sec_hewind}).
This may be a theoretical explanation for the different distributions of nova speed class between
different galaxies as introduced in Section \ref{sec_introduction}.

\section{NOVAE/SSS IN GLOBULAR CLUSTERS} \label{sec_GCnova}

As of July 2013, five novae and one nova candidate have been reported 
to be associated with globular clusters.
Table \ref{table_nova} gives a brief summary of these objects.
We obtained the absolute $V$ and $R$ band magnitudes of these objects as follows,
with $A_V=3.1~E(B-V)$ and $A_R=2.33~E(B-V)$ \citep{cal09},
where $A_R$ and $A_V$ are the Galactic extinction toward a cluster. 
The absorption free distance modulus is estimated to be 
$\mu_0=24.47 \pm 0.035 \pm 0.045$ from red clump stars \citep{sta98}, 
$24.47 \pm 0.07$ from tip of the red giant branch \citep{mcc05},
$24.36 \pm 0.11$ from eclipsing binaries \citep{vil10}, and  
$24.38 \pm 0.06 \pm 0.03$ from Cepheids \citep{rie12}. 
We adopt the arithmetic mean of these values, i.e., $\mu_0=24.42$ (distance of 766 kpc) in 
the present work. 

For Bol 126 (B126), we obtain $M_R=-7.89$ from $R=16.7$ with $A_R=0.17$ mag \citep{hen13}
and $M_V=-7.53$ from $V=17.12$ \citep{gal07}.
For Bol 111 (B111), we obtain $M_V=-8.36$ and $M_R=-8.65$ from $V=16.80$ and $R=16.33$ 
\citep{gal07} with $E(B-V)=0.24$ \citep{van10}.
For Bol 194 (B194), we obtain $M_V=-7.88$ and $M_R=-8.18$ from $V=17.19$ and $R=16.73$ 
\citep{gal07} with $E(B-V)=0.21$ \citep{van10}.
For other globular clusters, references are given in the table caption.
These magnitudes already appeared in Figures \ref{lightcurve} and \ref{nova.rmag}.
In what follows, we describe our light curve analyses for
the first three objects in Table \ref{table_nova}.

\subsection{M31N 2007-06b in Bol 111} \label{sec_M31N2007-06b}

This nova was discovered by Quimby on June 19.38 UT, 2007
at an unfiltered magnitude of $16.96 \pm 0.07$.
It reached $16.89 \pm 0.06$ on June 21.38 UT and had faded to below 17.89
by June 30.33 UT.
\citet{sha07} estimated the fading rate to be $> 0.11$ mag/day, suggesting a fast nova.
The spectrum taken roughly 3 days after discovery is characterized by
strong and broad (FWMH $\sim 3000$ km~s$^{-1}$) Balmer emission lines \citep{sha07}.
With the absence of significant \ion{Fe}{2} emission features, these authors
concluded that this nova is a member of the He/N class \citep{wil92}.


\begin{figure}
\epsscale{1.15}
\plotone{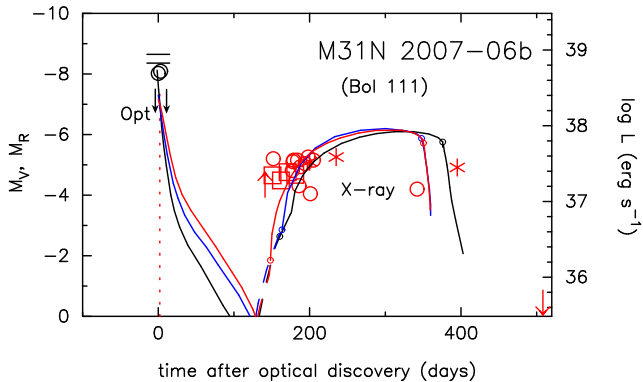}
\caption{Light curves of optical and supersoft X-ray of M31N 2007-06b.
Black open circles and downward arrows (upper limit) depict unfiltered optical magnitudes
taken from \citet{sha07}.
A red vertical dotted line indicates the suggested maximum (June 21.38 UT, 2007,
i.e., 2 days after the discovery).
X-ray data are taken from \citet{hen09} but the data are shifted by $-2.5$ dex,
in order to fit with the observed turn-on and turnoff times.
Different symbols indicate different satellites. Red open squares: Chandra HRC-I,
Red open circles: Swift XRT. Red asterisks: XMM-Newton EPIC pn.
An lower limit (red upward arrow) on day 152 is taken from \citet{hen09}, and
an upper limit (red downward arrow) at day 508 from \citet{hen11}.
Two short black horizontal lines indicate the brightness of Bol 111 in $R$ (upper) and $V$
(lower) magnitudes.
Solid lines depict the model light curves.  Black line: $(M_{\rm WD},~X,~X_{\rm O},~X_{\rm Ne},~Z)$
=(1.15 $M_\odot$, 0.55, 0.1, 0.03, 0.02).
Blue line: (1.18 $M_\odot$, 0.55, 0.1, 0.1, 0.004). 
Red line: (1.2 $M_\odot$, 0.65, 0.2, 0., 0.001). 
\label{SS1}
}
\end{figure}

The peak magnitude of $16.96 \pm 0.07$ is comparable to the magnitude of the host cluster
Bol 111.
Assuming the unfiltered magnitude to be close to the $R$ magnitude,
we estimated the peak absolute $R$ magnitude to be $M_R \sim -8.02$
with $\mu_0=24.42$ and $A_V=0.24 \times 2.33 = 0.56$.
There were only two upper limit observations
before and after the assumed optical peak (Figure \ref{SS1}).
One might think that this nova has multiple peaks like V723 Cas and V5558 Sgr and that
we missed the first peak.
This possibility is rejected because the short X-ray turn-on time of M31N 2007-06b
indicates that this nova is a fast nova that likely shows a smooth light curve with a single peak
and not a multi-peak slow nova like V723 Cas \citep[its turnoff time is
longer than 12 years:][]{nes08}.
Another possibility is a small group of novae that shows sharp secondary peak like
V1493 Aql and V2491 Cyg.
This case is also unlikely because the peak magnitude is too large as the secondary
peak and the pre-maximum upper limit may be inconsistent with a relatively slow rising of the
secondary peak.
Therefore, in the present work we regard M31N 2007-06b as a fast nova with a single maximum.

Figure \ref{SS1} depicts the X-ray light curve of M31N 2007-06b from the day of the discovery.
\citet{hen09} detected supersoft X-ray emission on day 141,
but found no detection on day 33, thus, the X-ray turn-on time is between both observations.
Also, the turnoff time is between day 395 and day 508 \citep{hen09,hen11}.
This relatively rapid appearance and disappearance of a supersoft X-ray
phase suggest a massive WD.
A massive WD is also suggested by the relatively high blackbody temperature of $48 \sim 70$ eV
which was estimated from the X-ray spectrum \citep{hen09}.

We have calculated many theoretical light curves from which
three models with different $Z$ are shown in Figure \ref{SS1}.
These models more or less show good agreement with the X-ray data.
If we decrease $Z$, we must increase the WD mass and $X_{\rm O}$ and/or $X_{\rm Ne}$
in order to fit the X-ray turn-on time. At the same time we need to increase $X$
to fit with the X-ray turnoff time. Therefore, we have a limited range within
the parameter set of the WD mass and composition.
The spectroscopic metallicity of Bol 111 is obtained as [Fe/H]$=-0.85~\pm 0.71$
\citep{gal09} and $-1.3 \pm 0.1$  \citep{cal11}. Consequently, we adopt a $Z=0.001$ model.
This model of $M_{\rm WD}=1.2 M_\odot$ indicates the photospheric temperature
rising from $\log T$(K)=5.37 (20 eV) at the X-ray turn-on up to $\log T$(K)=5.94 (74 eV)
just before the X-ray turnoff and about $\log T$(K)=5.78 (52 eV) at day 235.
\citet{hen09} estimated the temperature from the XMM-Newton EPIC pn X-ray spectra on day 200 and
235 to be $\sim 61 \pm 1$ eV, assuming a plane parallel model with a halo composition.
Considering the ambiguity of our model
(fitting ambiguity and adopted simplification) and X-ray analysis
(the plane parallel assumption is not good near the turn-on time because the
photospheric radius is large), the observed value is consistent with our theoretical estimate.

\subsection{M31N 2010-10f in Bol 126} \label{sec_M31N2010-10f}

M31N 2010-10f appeared in the M31 globular cluster Bol 126 \citep{cao12,hen13}
on MJD 55480.32.
\citet{hen13} estimated the $t_2$ time of the $R$ band light curve
as $t_{2,R} \leq 3.12$ days, suggesting a very fast nova.
The peak magnitude of the nova was estimated to be $R \sim 16.1 \pm 0.3$, by
subtracting the contribution of the globular cluster ($R=16.7$) from
the observed magnitude $R \sim 15.7$ at the peak.
The optical magnitude of Bol 126 was practically constant at $R=16.7 \pm 0.1$ mag
during the 74 days before the outburst. Therefore, we can safely exclude the possibility
that we missed other multi-peaks before the observed peak at MJD 55480.32.


\begin{figure}
\epsscale{1.15}
\plotone{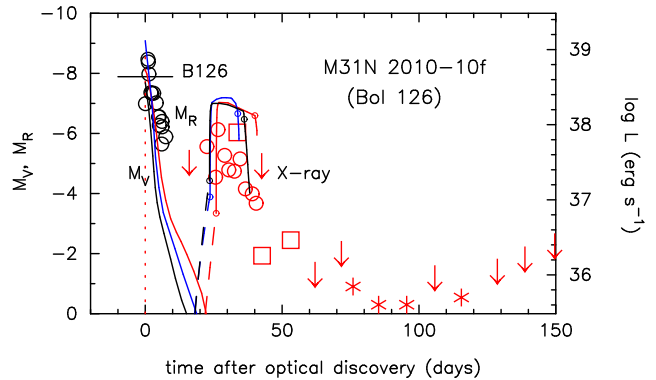}
\caption{Light curves of M31N 2010-10f in the optical $R$ band (black open circles)
and in the X-ray (red open circles, squares, and asterisks).
Data are taken from \citet{hen13}.
Solid lines in the left-hand side indicate theoretical optical light curves  that represent
the $V$ magnitude light curve, or more exactly the emission-line-free $y$ band.
The solid lines on the right-hand side indicate X-ray (0.2 -- 1.0 keV) light curves. 
Black lines: ($M_{\rm WD},~X,~X_{\rm C}$,~$X_{\rm O},~X_{\rm Ne}~,Z$)=
(1.36 $M_\odot$, 0.35, 0.1, 0.2, 0., 0.02).
Blue lines: (1.37 $M_\odot$, 0.5, 0., 0.2, 0.1, 0.004).
Red lines: (1.37 $M_\odot$, 0.35, 0., 0.2, 0.1, 0.001).  
Other symbols are the same as those in Figure \ref{SS1}. 
\label{novaB126}
}
\end{figure}

Figure \ref{novaB126} shows the $R$ band light curve of M31N 2010-10f.
X-rays were not found 16 days after the optical discovery, but detected
on day 22.5.
The nova shows a long X-ray tail after the X-ray luminosity dropped by a factor of 20
on day 40. We regard this point in time as the extinguishing point of the
hydrogen nuclear burning. 
Even after the nuclear burning has almost extinguished, the WD surface
is still hot and we expect that scattered X-ray continuum or reprocessed line emission contributes  
to the X-ray flux.
A similar tail after the X-ray turnoff was reported in the classical nova V2491 Cyg
\citep{hac09,pag10}.
The short X-ray turn-on time (day $16  < t_{\rm on} <$ day 22.5)
and a very short SSS phase ($\sim 18$ days) suggest a very massive WD \citep[e.g.][]{hac06}.

We calculated many light curve models with different compositions,
three of which are shown in Figures \ref{novaB126}.
The optical light curves represent emission-line-free $y$ band magnitudes, so we cannot
directly fit it with the $R$ band data because emission lines contribute to the
$R$ magnitude.
Therefore, we only adjust $t=0$ of the theoretical curve to the observed peak
magnitude but neglect the difference of decline rate between theoretical and
observed light curve. As in Figure \ref{nova.rmag}, light curves always decay slower in $R$
than in $V$. In this sense, our theoretical light curves are consistent with
the observed $R$ band data.
As the X-ray turn-on and turnoff time are very short, only very massive WD models
show a good fit.
The three models in this figure are
($M_{\rm WD},~X,~X_{\rm C}$,~$X_{\rm O},~X_{\rm Ne}~,Z$)=(1.36 $M_\odot$, 0.35, 0.1, 0.2, 0.0, 0.02),
(1.37 $M_\odot$, 0.5, 0.0, 0.2, 0.1, 0.004), and
(1.37 $M_\odot$, 0.35, 0.0, 0.2, 0.1, 0.001).
The chemical composition of X, $Y$, $X_{\rm C}$, $X_{\rm O}$, and $X_{\rm Ne}$
are not unique for a given $Z$.
For example, we can decrease/increase $X_{\rm Ne}$ while keeping $X_{\rm Ne}$ + $Y$ constant.
If we increase $X$, the light curve becomes slower, which can be compensated
by an increase of the WD mass or $X_{\rm O}$. However, the light curve evolves very
fast, so that we cannot reproduce it with low-mass WD models.
The metallicity of Bol 126 is spectroscopically estimated to be [Fe/H]=$-1.20 \pm 0.47$
\citep{bar00}, $-1.48 \pm 0.19$ \citep{gal09} and $-1.5 \pm 0.2$ \citep{cal11}.
The above case of $Z=0.02$ is too large, so that it may give an lower limit of the WD mass.
Therefore, we may conclude that the WD mass is as massive as $> 1.36~M_\odot$.

The photospheric temperature of the WD envelope increases with time
and reaches its maximum shortly before the X-ray turnoff. In the above models of 1.37 $M_\odot$
the temperature increases from  $\log T$ (K)= 5.43 at the X-ray turn-on
to 6.16 (i.e., from 23 to 130 eV) just before the turnoff time (right small open circle).
After that, the photospheric temperature decreases because nuclear burning
gradually extinguishes and internal heat energy is lost by radiation.
In the later phase of nova outbursts, emission lines dominate the optical light.
Our model does not include such contributions of lines, so we stop calculation shortly
after the turnoff time.

\citet{hen13} derived the blackbody temperature at the stage of days 76 -- 115
to be $kT = 74 ^{+30}_{-26}$ eV with $N_{\rm H} = 0.4 ^{+0.9}_{-0.4} \times 10^{21}$ cm$^{-2}$
from XMM-Newton spectra.
This epoch corresponds to the cooling phase after nuclear
burning extinguished. The temperature is consistent with our theoretical model.
The cooling timescale depends on the opacity, therefore, it depends on the heavy element
content of the envelope. The cooling process of a post-nova in a low $Z$ environment
has not been studied well.
Thus, these temperature estimates are important information for future study.

Almost all of the Galactic classical novae are short period binaries, typically, as short as
$P_{\rm orb}=$ a few to several hours.
\citet{hen13} examined X-ray variation and found no short-time modulation
less than 5.56 hr (the longest exposure time during the luminous SSS phase).
This indicates that M31N 2010-10f is not an eclipsing binary with orbital period of
$P_{\rm orb}~<~5.56$ hr.

Massive WDs have been suggested in some Galactic novae.
The fast nova V2491 Cyg shows a short SSS phase of only 10 days starting from
day 38 after outburst. Such a short duration of the SSS phase is interpreted
as indicating a WD of about 1.3 $M_\odot$ \citep{hac09}.
Similar WD masses of $\sim 1.3 M_\odot$ are also suggested
for the classical novae V693 CrA \citep{kat07h} and V377 Sct \citep{hac07}.
In the very fast classical nova V838 Her, the WD mass was estimated to be $1.35 \pm 0.02 M_\odot$
\citep{kat09}. Extremely massive WDs ($\sim 1.37 M_\odot$)
are suggested in recurrent novae and type Ia supernova (SN) progenitors
\citep[e.g.][for summary]{kat12review}.
It would be interesting to know whether M31N 2010-10f is a recurrent nova or not.
However, we do not have sufficient data to support short recurrence times.

\subsection{Transient SSS in BOL 194: A Nova Candidate} \label{sec_SSS_Bol194}

\citet{hen09} reported a transient supersoft X-ray source which lasted about one month.
This object is a nova candidate, from its X-ray spectrum and duration,
but no optical counterpart was detected
in the previous 1100 days \citep{hen09}. A possible interpretation is that
an outburst occurred during several seasonal gaps of unobservable
periods, i.e., 257 -- 166, 621 -- 542, and 990 -- 904 days before
the first positive X-ray detection \citep{hen09}.
The other possible interpretation is that the nova was much fainter
than the host cluster Bol 194.


\begin{figure}
\epsscale{1.15}
\plotone{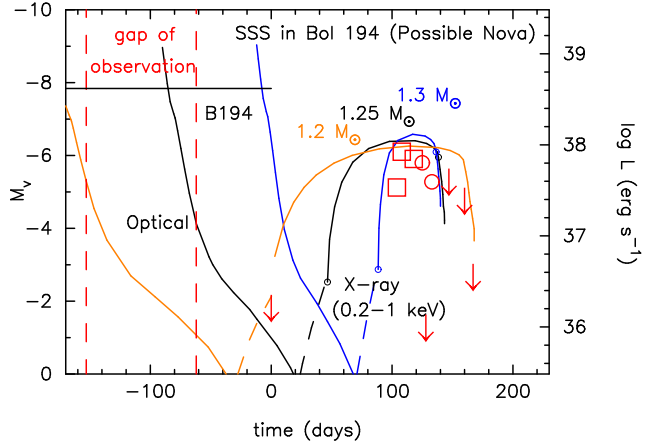}
\caption{Unabsorbed X-ray light curve of the transient supersoft X-ray source in Bol 194,
CXO J004345 \citep[=SS2 in][]{hen09}.
The X-ray phase began between day 0 and 100 and ends about day 130 in this
figure. Data are taken from \citet{hen09}.
Three theoretical models of $M_{\rm WD}=$ 1.2,~1.25 and 1.3 $M_\odot$ WDs
($X=0.55,~X_{\rm O}=0.1,~X_{\rm Ne}=0.1,~Z=0.001$) are shown for both optical
and supersoft X-ray bands.
A seasonal gap of observed period is shown by the two dashed lines.
The horizontal solid line indicates the absolute $V$ magnitude of Bol 194.
\label{SS2}
}
\end{figure}

Figure \ref{SS2} shows our theoretical light curve model of a 1.25 $M_\odot$ WD
as well as two other models of 1.2 $M_\odot$ and 1.3 $M_\odot$ WDs.
Here, we adopt $Z=0.001$ after the spectroscopic abundance of Bol 194, [Fe/H]=$-1.4 \pm 0.1$,
obtained by \citet{cal11}. All of these models show a supersoft X-ray phase (0.2 -- 1.0 keV)
more or less consistent with the data reported by \citet{hen09},
but only for the 1.25 $M_\odot$ WD model the optical peak
fell in the seasonal gap.
Note that these theoretical curves begin from a very bright peak ($M_V \sim -9$),
but the actual optical peak is different from object to
object, because it is determined by the ignition mass (i.e, the mass-accretion rate).
The estimated WD mass (1.25 $M_\odot$) is so high that we expect the peak magnitude to be
similar to a typical Galactic fast nova, $M_{\rm V} < -8$ (see Figure \ref{lightcurve}).
Thus, the outburst could have been detectable if it had occurred in the observable period.

There are other observational gaps of 621 -- 542 and 990 -- 904 days before the X-ray detection.
If the nova had erupted during these periods, the turn-on and turnoff times would become
much longer with the duration of the supersoft X-ray phase being unchanged.
We cannot reproduce such long turn-on/off times but short X-ray duration
simultaneously, even if we assume a low mass WD with different set of chemical composition.
Therefore, we can reject the possibility of a nova outburst in the previous gaps.
In this way, we may conclude that only the 1.25 $M_\odot$ WD model is consistent with
the observations.

The WD temperature of the 1.25 $M_\odot$ WD model is $\log T$ (K) $\sim 5.98$ (82 eV)
at the time of the X-ray observation with Swift. This value is consistent with
the temperature $kT = 74^{+32}_{-23}$ eV
deduced from a Swift XRT X-ray spectrum \citep{hen09},
although the spectrum is very poor and the error range is quite large.

Therefore, we conclude that this transient supersoft X-ray source CXO J004345 in Bol 194
\citep[=SS2 in][]{hen09} is highly likely a nova remnant.


\begin{figure}
\epsscale{1.15}
\plotone{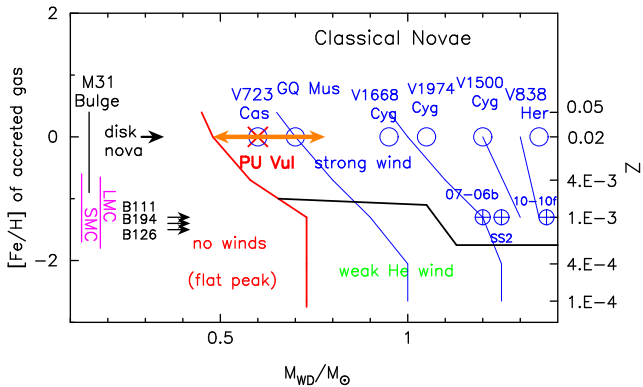}
\caption{Similar to Figure \ref{MwdZ} but we added individual novae.
Upper right blue symbol region above the black and red lines:
strong winds driven by the iron OPAL peak at $\log T$(K) $\sim$ 5.2.
Lower right region below the black line: possible weak winds driven by the He o\
pacity peak.
The region left-hand-side to the red line: no winds are accelerated.
Three novae are shown by open circles with a plus inside. SS2 indicates the
transient supersoft X-ray source in Bol 194 \citep{hen09}.
07-06b: M31N 07-06b, 10-10f: M31N 10-10f. See text for more details.
\label{MwdZ.obj}
}
\end{figure}

\subsection{Summary of Novae in Globular Clusters}\label{sec_nova_summary}
  
Figure \ref{MwdZ.obj} shows various novae in the $M_{\rm WD}$ -- $Z$ diagram.
This is a similar diagram to Figure \ref{MwdZ} but
we added Galactic novae PU Vul, V723 Cas, GQ Mus, V1668 Cyg, V1974 Cyg, V1500 Cyg, 
and V838 Her at $Z=0.02$.
We adopt the WD mass of GQ Mus to be $M_{\rm WD}=0.7 \pm 0.05 M_\odot$ \citep{hac08}
and of V1974 Cyg to be $M_{\rm WD}=1.05 \pm 0.05 M_\odot$ \citep{hac06}.
For the other novae see Section \ref{sec_galacticnova}.
These are well observed classical novae and their WD masses are estimated
from theoretical light curve fittings using multiwavelength observations including 
IR, optical, UV, and supersoft X-ray.


These Galactic disk novae are distributed between 0.6 $M_\odot$ and 1.35 $M_\odot$ on the
line of $Z=0.02$. In general, a more massive WD corresponds to
a faster nova and a less massive WD to a slower nova.
As described in Section \ref{theory_Zdepencence}
strong optically thick winds are accelerated in the blue region
and no optically thick winds are accelerated in the region to the left of the red line, thus
those novae evolve extremely slowly.
In a limited mass range indicated by an orange two-headed arrow at $Z=0.02$
(from about 0.5 to 0.8 $M_\odot$, which depends on the chemical composition),
both the static and wind evolutions are realized.
For example, the classical nova GQ Mus underwent the wind evolution from the
beginning of the outburst as in the other massive classical novae \citep{hac08}. 
The symbiotic novae PU Vul underwent the static evolution without optically thick winds 
\citep{kat11a,kat12hm}. 
The slow novae V723 Cas, HR Del and V5558 Sgr are the transition novae,
which started with the static evolution but halfway made a transition to the
wind evolution \citep{kat11h}. A transition nova shows multiple peaks
and is fainter than the other fast novae as shown in Figure \ref{lightcurve}.

These properties are known in Galactic novae but have not been studied in low metallicity novae.
We added three novae in globular clusters studied in the present work.
All of them have a massive WD, thus they locate in the
lower-right region ($Z \sim 10^{-3}$ and $M_{\rm WD}=$ 1.2 --1.37 $M_\odot$)
of the blue part. None of them is found in the green part
where the He opacity peak is responsible for the acceleration.
The present work is the first step to understand the characteristics of
low metallicity novae in the $M_{\rm WD}$ -- $Z$ diagram.

\section{Discussion} \label{sec_discussion}

\subsection{Importance of X-ray detection of Novae in Globular Clusters} \label{discussion_Xray}

In Galactic novae, supersoft X-ray phases are detected in a number of objects and
are used, together with other multi-wavelength light curves, to estimate the WD mass.
These WD masses range from $0.6$ to~1.37 $M_\odot$
\citep{hac06,hac07,hac09,hac10,hac07l,hac08,kat11a,kat09,kat12review}.
Such information is crucial to understand the physics of nova outbursts.
Similarly, X-ray observations are also crucial for quantitative
studies of extragalactic novae.
In case of extragalactic novae, or Galactic bulge/halo novae, determining the metallicity and
accurately detecting the X-ray turn-on and turnoff times are important.
In this work, we showed that even if no optical information is available
we can roughly estimate the WD mass from X-ray light curves with estimated metallicity
(see Section \ref{sec_SSS_Bol194} and Figure \ref{SS2}).

Optical surveys tend to detect relatively bright novae on massive WDs
because of the contamination of the nova signal by the host clusters. This limitation is much relaxed
in the case of X-ray surveys.
Galactic globular clusters locate far above the Galactic plane and M31 globular
clusters are found apart from the M31 disk. Therefore, we can expect a relatively small foreground 
absorption
in addition to a negligibly small internal extinction within each cluster.
If we choose a cluster of low extinction, we may detect the X-ray phases of novae
on relatively low mass WDs
even in very bright globular clusters for which an optical nova detection is difficult
(see Figure \ref{nova.rmag}).


\begin{figure}
\epsscale{1.15}
\plotone{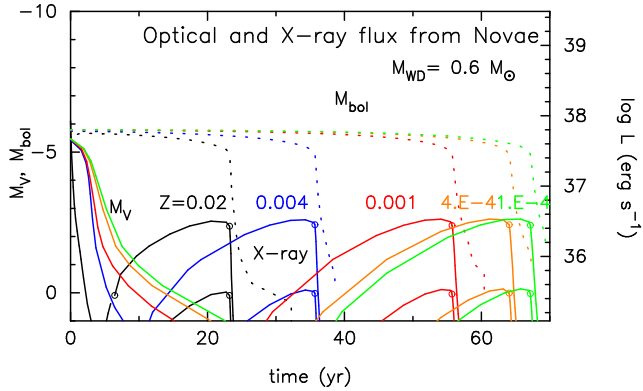}
\caption{
Theoretical light curve models of novae on 0.6 $M_\odot$ WDs
with $Z=0.02,~0.004,~0.001,~0.0004,$ and 0.0001.
Each model shows the bolometric magnitude (dotted lines), optical light curve (left solid lines),
and supersoft X-ray light curve of 0.2 -- 1.0 keV (12.4 to 62 \AA) (upper solid lines), and
0.3 -- 1.0 keV (12.4 -- 41.32 \AA) (lower solid lines).
Optical light curves are free-free emission for $Z=0.02$ model
and the photospheric blackbody emission for the other model because
optically thick winds do not occur.
Hydrogen nuclear burning extinguishes at the point marked by a small open circle.
All fluxes are unabsorbed. See text for more details.
\label{lightM06}
}
\end{figure}


\begin{figure}
\epsscale{1.15}
\plotone{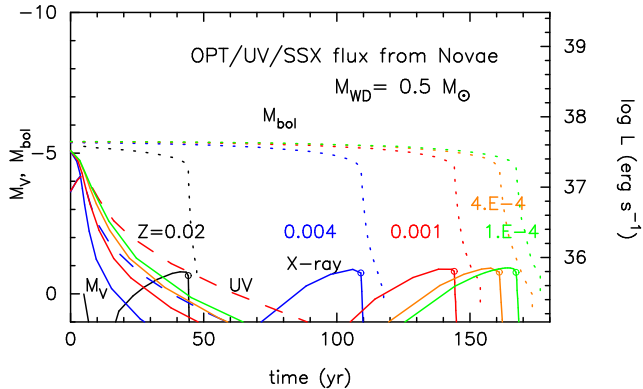}
\caption{Same as Figure \ref{lightM06}, but for 0.5 $M_\odot$ WDs.
UV light curves of the $2316 \pm 530$ \AA~ (from 1786 \AA~to 2846 \AA) band (dashed lines)
corresponding to the GALEX window are added for the two models of $Z=0.001$ and 0.004.
All the optical light curves begin with $T_{\rm ph}=10^4$ K at $t=0$.
The optically thick winds do not occur except for $Z=0.02$. In the case of $Z=0.02$,
only a later part is shown here owing to numerical difficulty.
The light curves are calculated using blackbody emission of the photospheric temperature
except for the optical free-free light curve of $Z=0.02$ model.
\label{lightM05}
}
\end{figure}


\begin{figure}
\epsscale{1.15}
\plotone{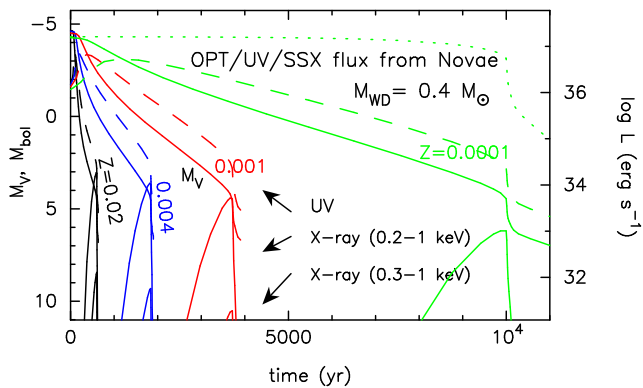}
\caption{
Same as Figure \ref{lightM06}, but for 0.4 $M_\odot$ WDs.
Lines for each model denote, from upper to lower,
UV light curve (dashed lines: corresponding to the GALEX window),
the $V$ band light curves (upper solid lines), and supersoft X-rays (lower solid lines).
The uppermost dotted line denotes the bolometric magnitude of the $Z=0.0001$ model.
All of these light curves are calculated from blackbody emission of the photospheric
temperature because no optically thick winds are accelerated.
\label{lightM04}
}
\end{figure}

Figure \ref{lightM06} shows unabsorbed model X-ray light curves (0.2 --1.0 keV for the upper line and
0.3 -- 1.0 keV for the lower line)
for $0.6~M_\odot$ WDs with various metallicities. The X-ray peak flux is almost independent
of the metallicity $Z$, but turn-on and turnoff times increase with decreasing $Z$.
The X-ray fluxes are much weaker for the 0.5 $M_\odot$ and 0.4 $M_\odot$ WDs
(see Figures \ref{lightM05} and \ref{lightM04}).
These X-rays are weak but detectable with the current satellites if the nova appears
in globular clusters of our Galaxy or in the LMC/SMC.
In case of M\,31, while the $0.6~M_\odot$ WDs might still be detectable in a long X-ray monitoring survey,
like the one described in \citet{hen11}, there is little possibility of detecting 0.5 $M_\odot$ or even
0.4 $M_\odot$ WDs using the current instruments.

In low metallicity novae, the X-ray turn-on started long after the optical outburst,
which makes it difficult to find its optical counterpart.
Long turn-on and turnoff times are observed in Galactic novae,
$t_{\rm on}\sim 400$ days for V458 Vul \citep{nes09},
$t_{\rm off}\sim$ 10 years for GQ Mus \citep{sha95a,ori01},
and $t_{\rm off} >$ 12 years for V723~Cas \citep{nes08}.
In a low metallicity environment, we expect much longer $t_{\rm on}$ and $t_{\rm off}$ times.
The metallicities of M31 novae are not known, but some of them may correspond to low
metallicity and/or low mass WDs.
The longest turn-on times yet observed in M31 novae are $t_{\rm on}\sim 13.8$ years for
M31N~1996-08b, and 11.5 years for M31N~1997-11a (Henze et al. 2013, in preparation).

Theoretically, all WDs in novae have a phase of stable nuclear
burning some time after the optical outburst.
\citet{hen11} showed for a relatively large sample of M31 novae that
the turn-on and turnoff times of the SSS phase are correlated with
each other as well as with the optical $t_2$ time.
These authors found there is no strong evidence for a large fraction
of self-absorbed novae, i.e. no objects for which the stable nuclear burning
ends before the ejected envelope turns optically thin.

This property is theoretically explained as follows. During the optical outburst,
winds are strong enough to blow most of the envelope mass far away from the
binary. The strong (optically thick) winds stop when
the photospheric temperature increases to beyond the Fe opacity peak
at $\log T$ (K) = 5.2 as shown in Figure \ref{opac}.
Thus, the X-ray turn-on time is almost coincident with the epoch at which the optically
thick winds stop (left-side open circles in Figures \ref{SS1}, \ref{novaB126}, and \ref{SS2}).
This enables us to detect supersoft X-rays from the hot WD without an obscuration by the ejecta.

In the very low [Fe/H] case, however, the optically thick winds may not blow at all or do so very weakly.
It is uncertain whether all of the ejecta will be quickly blown off
by the time of the X-ray turn on, as will be explained in Section \ref{sec_hewind}.
If the ejecta are not quickly blown off but stay around the orbital plane
until the X-ray turnoff time,
we may not always detect strong supersoft X-rays from the hot WD.
Therefore, the detection/non-detection of X-rays provide us valuable information
on the wind mass loss in nova outbursts.

\subsection {On the Optical Observation} \label{discussion_optical}

In spite of the contamination by the light of the host globular clusters,
in the optical we can detect nova outbursts of
similar magnitude to the globular cluster or even fainter.
For example, M31N 2007-06b was detected even though the peak magnitude
was fainter than that of the host cluster Bol 111 (see Figure \ref{SS1}).
\citet{hen13} deduced an optical light curve of M31N 2010-10f until it had decayed by more than
two magnitude below the level of the host cluster (Figure \ref{novaB126}).
This is possible because a nova outburst is the most luminous phenomenon in a globular cluster 
except of supernova explosions. The large increase in flux from the nova leads to a brightening of the
globular cluster. This increase is significant with respect to the
intrinsic uncertainty of the cluster magnitude without the nova outburst.
Thus, we typically can detect nova outbursts which are 1 -- 2 mag fainter than the cluster
with $3\sigma$ confidence even using small telescopes.
Large telescopes and long exposures such as WeCAPP survey
\citep[The Wendelstein Calar Alto Pixellensing Project:][]{rif01} and
PAndromeda \citep[Pan-STARRS 1 survey of M31:][]{pandro12}
push the detection limit towards much fainter and slower novae.
The detection of the optical peak is very useful
in estimating the WD mass together with the X-ray data, as shown in Section \ref{sec_SSS_Bol194}.

Figure \ref{MwdZ} shows a systematic difference in nova speed class between
globular cluster novae (typically $Z\sim 0.001$) and Galactic disk novae ($Z= 0.02$).
For the same WD mass, novae evolve much slower in globular clusters than
in the Galactic disk.
At the present stage, we cannot confirm our prediction in Figures \ref{fast} and \ref{slow},
a systematic delay of supersoft X-ray phase with increasing $Z$ for the same
optical decay rate. Optical magnitudes of novae studied here are obtained
without filter (M31N 2007-06b) or in the
$R$ band (M31N 2010-10f), but these photometric bands do not represent the photospheric evolution.
The most appropriate way to monitor the evolution of the
visual continuum is using the Str\"omgren $y$ filter, which is designed to
avoid strong emission lines, in particular, [\ion{O}{3}] $\lambda\lambda 5007$ and
4959 lines.
The second appropriate way is to use the $V$ filter because most novae
show a similar magnitude decay in $V$ and $y$ bands
until a few magnitude below the optical peak.
Our light curve model does not include line formation, which is object dependent,
in order to represent common physics of nova outbursts.
The $R$ band magnitudes are contaminated by many emission lines, including most prominently H$_\alpha $, 
and
yield entirely different light curves from $V$ as shown in Figure \ref{nova.rmag}.
The $y$ band is moderately narrow, thus it may not be suitable to follow extragalactic novae,
because we need to observe the optical light curve until it becomes fainter 
by at least 2 -- 3 magnitude from the optical peak or longer if possible.
Therefore, we encourage the use of $V$ filter observations instead of, or together with,  
the $R$ band, or without any filter.

Novae show a wide variety in their light curves, lines, colors, dust, evolution timescales, etc.,
but there are common properties that reflect the fundamental physical process of the nova outburst.
One of the well-known properties is the relation between the
maximum magnitude and rate of decline of a nova light curve (MMRD relation).
There are many MMRD relations proposed and the general trend is theoretically explained by
\citet[][also for review of MMRDs]{hac10}.
Other common properties include the universal decline law of nova light curves \citep{hac06}
and a common path of evolution in the color-color diagram \citep[in (B-V) -- (U-B) diagram:][]{hac13}.
The MMRD relation was established for novae discovered in our Galaxy, M31 and the LMC.
Thus, this relation is common among various metallicities.
The universal decline law and color evolution are obtained from the analysis of well observed
Galactic novae and have been applied to a number of other Galactic novae, but
not yet exactly confirmed in extremely low metallicity environments.
Recently, \citet{lee12} applied the universal decline law to
M31 novae detected as a by-product of searching for microlensing events
by the WeCAPP project.
The decline rates of these nova light curves are consistent
with the theoretical predictions, considering the fact that their $R$ light curves are
contaminated by emission lines and the possibility
that they missed some of the optical peaks.
However, the metallicity is not known for these novae and we have not yet confirmed whether
the universal decline law exactly holds for low metallicity novae.
A common path in the color-color diagram of novae has been just discovered
but has not been examined for extragalactic novae.
Thus, novae in globular clusters are an ideal laboratory to extend our knowledge
of nova physics to very low metallicity environments.
High cadence light curves and color information will open the quantitative study of
extragalactic novae.

\subsection {Novae with Helium accelerated winds} \label{sec_hewind}

In low metallicity novae, we expect only weak wind acceleration owing to the helium opacity peak
(see Figure \ref{opac}). Such a helium acceleration wind occurs in the
green part of Figure \ref{MwdZ}.
We expect that $\sim 1.0~M_\odot$ WDs show slow nova outbursts, like a Galactic nova on a 0.7 $M_\odot$ WD,
as indicated by the solid line for slow novae in Figures \ref{MwdZ} and \ref{MwdZ.obj}.
However, their characteristic properties are highly uncertain.
All of our three novae studied in the present work, unfortunately, are found in
the blue part and we have no corresponding objects identified in the green region. 

Figure \ref{struc} shows the internal structure of a wind solution for a
1.0 $M_\odot$ WD and $Z=0.0004$, which is a typical solution of helium acceleration
that locates in the middle of the weak wind region in Figures \ref{MwdZ} and \ref{MwdZ.obj}.
The wind acceleration is weak and the expansion velocity barely reaches the
escape velocity of the WD.
In case of 0.9 $M_\odot$  ($Z=0.001$) the wind velocity does not exceed the escape velocity.
Figure \ref{struc} shows  that
the critical point (the center of the accelerating region) appears at $2.83~R_\odot$,
which is sufficiently outside the companion orbit in typical close binaries.
In both of the 1.0 $M_\odot$ WDs with $Z=0.0004$ and 0.0001, the critical point always locates
outside the orbit of typical close binaries throughout the outburst.

Such a wind solution may be realized if the outburst occurs in a long period binary like PU Vul
($P_{\rm orb}=13.4$ yr),
because the photosphere does not reach the companion even at the maximum expansion
so that the companion does not affect nova evolution.
In case of a close binary, however, the companion's gravity may decelerate the wind 
acceleration and
thus the nova outburst may be different from a normal Galactic slow nova.

Effects of the companion on the nova envelopes are theoretically studied
by \citet{kat91a,kat91b,kat94h,kat11h}.
In Galactic novae, the wind acceleration is strong enough and the
presence of a companion does not affect the nova light curve even when the companion is
engulfed deep in the envelope.
This is because the critical point locates inside the orbit and the wind is
already accelerated to a large velocity at the orbit.
Moreover, the density at the orbit is low enough and
frictional effects due to companion's motion do not produce sufficient additional energy
to affect the nova evolution.
On the other hand, when the acceleration is weak, effects of the companion's gravity
surpass the other effects, which suppresses the wind acceleration \citep{kat91a,kat91b,kat11h}.
It is highly uncertain, though, in such a case if
the ejecta could form a slowly expanding envelope around the orbital plane.
Thus, a nova outburst may not look like a normal Galactic nova, but a
multi-peak slow nova with occasional mass ejection,
or it might appear like a red giant variable with a slow mass-ejection.
If these novae undergo a very slow evolution with a timescale longer than 10 years,
they are hardly likely identified as a nova outburst.
If normal slow novae can be expected only in long period binaries
i.e., symbiotic novae, they are rather rare objects from the analogy of Galactic novae.
Thus, we may not expect a large number of nova outbursts in the middle green region.

\subsection {Novae in low metallicity environment}

Our present work predicts that novae in a lower
metallicity environment should be generally slower than novae in 
predominately Pop I environments. 
This is consistent with the statistical properties of nova populations 
reviewed in \citet[][Fig. 6]{sha13b}, where M31 novae appear to 
show on average a slower decline than their Galactic counterparts. 
This effect could be explained by observations which suggest that 
the majority of M31 novae belong to a 
bulge population \citep[e.g.][]{sha01} which has a slightly lower 
metallicity, as shown in Section \ref{theory_Zdepencence}. 
We also suggest that in a very low metallicity environment less massive WDs do not 
have normal nova outbursts because they do not develop strong winds (Section \ref{sec_hewind}). 
This is consistent with the fact that all the globular cluster novae 
are fast novae, in addition to the other possible explanation of contamination by the host 
cluster.  

\citet{sha13a} showed that LMC novae are faster on average than their Galactic counterparts.
Although the LMC nova population is not the subject of this paper, 
a tentative application of our models to this situation suggests 
that slow novae are not prevalent in this galaxy because low-mass WDs
are unable to drive nova winds in the LMC's low metallicity environment,
and therefore do not contribute significantly to the observed nova outbursts.
The referee pointed out, if only a limited mass range of WDs were 
to contribute to the nova population, one might expect that the 
luminosity-specific nova rate should be diminished
in LMC relative to other galaxies, 
which is contrary to what is observed \citep[e.g.,][]{del94,gut10,sha13b}. 
Again, we do not model the LMC nova rate here but suggest that
the LMC's relatively high luminosity-specific nova rate results
from a large population of
massive WDs formed in young star-forming regions of the galaxy.
Not only do massive WDs produce fast nova outbursts, they produce outbursts more
frequently than do less massive WDs because of their smaller ignition masses for
the same mass-accretion rate \citep[e.g.][]{pri95}. 
Thus, a small difference in the WD mass distribution will have a strong
effect on the overall nova rate. 
Finally, we note that young binaries are also expected to have more massive
companions that can drive higher mass-accretion rates and produce shorter
recurrence times. In summary, LMC novae produced in star forming regions are
expected to harbor more massive WDs and companion stars, both of which should
act to increase the specific nova rate.

\section{CONCLUSIONS} \label{sec_conclusions}

Our main results are summarized as follows.   

\begin{enumerate}

\item
We presented the first light curve analysis of novae in globular clusters and estimated the WD masses
to be $\sim 1.2~M_\odot$ for M31N 2007-06b in Bol 111 and
$\sim 1.37~M_\odot$ for M31N 2010-10f in Bol 126.
The WD temperatures based on these models are quite consistent
with those estimated from the X-ray spectra.
The transient supersoft X-ray source CXO J004345 in Bol 194 is consistent with
a nova model on a 1.2 -- 1.3  $M_\odot$ WD in its optical and X-ray light curves
and X-ray temperature. Thus, we regard it as a nova.

\item We presented theoretical light curve models for various metallicities of $Z$ ranging
from 0.05 to 0.0001. Novae in a lower metallicity environment evolve systematically
slower compared with Galactic disk novae ($Z=0.02$).
Assuming the WD mass distribution is the same,
we expect a tendency that the M31 bulge ($Z < 0.02$ for 60\% of stars)
hosts fewer fast novae and more slow novae
than the Galactic disk.

\item 
We predict a systematic difference between globular clusters of relatively
high ($Z > 0.001$) and low ($Z \lesssim 0.001$) metallicity.
In globular clusters of relatively high metallicity, various speed classes of novae may be detected.
In globular clusters of low metallicity, on the other hand,
a substantial part of novae may be slow and faint, and only fast novae
on massive WDs are prominent and detectable.
This is consistent with the fact that all the novae detected in globular clusters
so far are fast novae.

\item 
In optical surveys, we may preferentially detect
faster novae on massive WDs in bright clusters.
The detection limit depends on the cluster brightness,
metallicity, telescope aperture, and observing time.
On the other hand, in X-ray surveys, we are able to detect all the novae
on various WD masses except in very low metallicity clusters where circumbinary matter
may obscure low mass WDs.

\item
We encourage optical observation with the $V$ (or $y$) filter rather than $R$ to avoid a
contamination by emission lines and to secure light curve data
until at least two magnitudes below the optical peak.
Also, multi-wavelength observations including optical, UV, and supersoft X-ray are very important
in order to open the quantitative study of Population II novae.
\end{enumerate}

\acknowledgments

We are grateful to Akira Arai for providing information on M31 globular cluster nova observation
and also valuable discussion on multi-peak novae.
We are also grateful to Daisaku Nogami for providing us his new estimates 
of H$_\alpha$/continuum ratios of slow novae.  
We wish to thank the anonymous referee for useful comments that improved the manuscript.  
We also thank the American Association
of Variable Star Observers (AAVSO) for the visual data of
\object{V1668 Cyg}, \object{V723 Cas}, and \object{V5558 Sgr}.
This research was supported in part by the
Grants-in-Aid for Scientific Research (22540254 and 24540227)
from the Japan Society for the Promotion of Science.
MH acknowledges support from an ESA fellowship.

\end{document}